\documentclass[final, a4paper]{raa}            % referee version: for submission
\usepackage{graphicx,times} 

\begin{document}

\title{External Calibrator in Global Signal Experiment for Detection of  
the Epoch of Reionization}

   \volnopage{Vol.0 (200x) No.0, 000--000}      %%preserved for Editor. DOn't remove!
   \setcounter{page}{1}          %%starting page, preserved for Editor. DOn't remove!

\author{Yan Huang \inst{1}
       \and Xiang-Ping Wu \inst{1,2}
       \and Quan Guo \inst{2}  
       \and Qian Zheng \inst{2}         
       \and Biying Li \inst{1} 
       \and Huanyuan Shan \inst{2} 
       \and Kejia Lee \inst{1,3}
       \and Haiguang Xu \inst{4}
}

\institute{National Astronomical Observatories, Chinese Academy of Sciences, 
20A Datun Road, Beijing 100101, China; \\
\and
Shanghai Astronomical Observatory, Chinese Academy of Sciences, 
80 Nandan Road, Shanghai 200030, China \\
\and
Kavli Institute for Astronomy and Astrophysics, Peking University, 
5 Yiheyuan Road, Beijing 100871, China \\
\and
School of Physics and Astronomy, Shanghai Jiao Tong University, 
800 Dongchuan Road, Shanghai 200240, China \\
}

 \date{Received~~2020 month day; accepted~~2021~~month day}

\abstract{
We present a conceptual design study of external calibrators in 
the 21 cm experiment towards detecting the globally averaged radiation of
the epoch of reionization (EoR). Employment of external calibrator 
instead of internal calibrator commonly used in current EoR experiments 
allows to remove instrumental effects such as beam pattern, receiver 
gain and instability of the system if the conventional three-position
switch measurements are implemented in a short time interval. 
Furthermore, in the new design the antenna system is placed in 
an underground anechoic chamber with an open/closing
ceiling to maximally reduce the environmental effect such as RFI and ground 
radiation/reflection. It appears that three of the four external 
calibrators proposed in this paper, including two indoor artificial 
transmitters and one outdoor celestial radiation (the Galactic polarization), 
fail to meet our purpose. 
Diurnal motion of the Galactic diffuse emission turns to be the most possible 
source as an external calibrator, for which we have discussed the observational 
strategy and the algorithm of extracting the EoR signal. }


\authorrunning{Y. Huang, X.-P. Wu, Q. Guo, Q. Zheng, B. Li, H. Shan, K. Lee 
               \& H. Xu} 

\titlerunning{External Calibrator in EoR Experiment}

\maketitle

\section{Introduction}
\label{sect:intro}

Observational campaign for detection of the redshifted 21 cm line of 
neutral hydrogen from the dark ages, cosmic dawn (CD) and epoch of 
reionization (EoR) has entered into a golden era, with the recent detection 
of the prominent absorption feature around 70 MHz by EDGES experiment 
over the High-Band spectrum (Bowman et al. \cite{Bowman18}). 
Both the large depth and 
flat-bottomed shape of this signature are incompatible with the predictions 
of standard cosmological model of CD and EoR. This 
has triggered many discussions and speculations including 
exotic models of dark matter interaction 
(e.g. Barkana \cite{Barkana18}; 
Fialkov et al. \cite{Fialkov18}) and 
an excess radiation background above the cosmic microwave radiation   
(e.g. Feng \& Holder \cite{Feng18};  
Ewall-Wice et al. \cite{EwallWice18} ; 
Fialkov \& Barkana \cite{Fialkov19}).  
Further observations will absolutely be needed to confirm this finding. 

All the ongoing experiments towards the measurement of the globally 
averaged radiation of EoR (hereafter EoR experiment)   such as 
EDGES (Bowman \& Rogers \cite{Bowman10}; 
      Bowman et al. \cite{Bowman08}), 
BIGHORNS (Sokolowski et al. \cite{Sokolowski15}),
SCI-HI (Voytek et al. \cite{Voytek14}), 
LEDA (Price et al. \cite{Price18}), 
SARAS (Patra  et al. \cite{Patra13}), and 
PRI$^Z$M (Philip et al. \cite{Philip19}) 
are all based on the dipole-type antennas and their variants. 
Indeed, the choice of dipole-type antennas reduces the complexity of 
antenna system, leaving the major tasks to be calibration  
of the whole receive system and removal of extremely 
bright foreground, the two key challenges in the EoR experiment. 
There have also been proposals for detection of global CD/EoR signatures 
with short-spacing interferometers if some dedicated configurations 
are employed (Presley et al. \cite{Presley15};  
Singh et al. \cite{Singh15}). Otherwise, lunar occultation may offer
another opportunity for interferometers to achieve the goal 
(Mckinley et al. \cite{Mckinley18}).

Calibration in current EoR experiments is essentially accomplished 
by the conventional three-position switch measurements by connecting 
receiver to each of the following three inputs: 
the ambient load, an internal noise source 
as calibrator, and the antenna, a process called internal calibration 
which is fully and easily controlled (see Singh et al 2018 for a summary). 
This allows the sky temperature to be observationally determined 
provided that the accurate knowledge of antenna response is a priori known  
and the whole system remains stable throughout the measurements. 
While receiving system in each of the EoR experiments is mounted on a metal 
ground plane to reduce the ground radiation and reflection
(but see Bradley et al. \cite{Bradley19}), the antenna is actually exposed to 
environment noise despite that site is usually chosen to be remote 
and radio quiet. 

External calibration in the EoR experiment seeks for a spectrally smooth and 
broadband signal external to the receiving system that has the same 
radiation path and sky coverage as those of the background cosmic signal. 
Here we have added the sky coverage as additional constraint to 
distinguish the traditional individual standard sources as calibrators 
such as Cas A. This may allow us to avoid the beam correction in 
the EoR experiments, helping keep the whole system completely free of  
systematics. Unfortunately, none of the ongoing EoR experiments has 
actually used  the external calibration. 
EDGES has once tested the diurnal motion of the 
Galactic emission as an external calibrator but found that it is very 
sensitive to the beam correction 
(EDGES MEMO \#215\footnote
{http://www.haystack.mit.edu/ast/arrays/Edges/EDGES-memos/}). 
A drone-mounted calibrator was employed for HERA, 
which indeed reached the desired 
precision but there are some systematic errors and uncertainties that 
need to overcome (Jacobs et al. \cite{Jacobs16}). 

In this paper we explore a novel conceptual design of the external 
calibrator for the 21 cm EoR experiment, based on an underground anechoic 
chamber at low frequency to create a radio-quiet space for calibrating 
the system and meanwhile isolating the environment and ground interference.
Both indoor calibrators using artificial transmitters and outdoor 
calibrators relying on celestial sources will be considered. We wish to 
design a novel EoR experiment system in terms of both theoretical 
constraints and engineering feasibility.

\section{Indoor calibration: a deep anechoic chamber} 
\label{sect: indoor1}

We begin with an underground anechoic chamber at low 
frequency to demonstrate our conceptual design. A schematic of the anechoic 
chamber is shown in Fig.1, which is a circular shaft with an opening/closing
ceiling, designed to absorb ambient background electromagnetic 
radiation. The walls, ceiling and floor of the anechoic chamber are treated
with wedge-shaped absorbers and further 
shielded by metal mesh to prevent radiation from ground.
A linearly-polarized dipole antenna, optimized to operate at 
frequencies between 50 MHz and 200 MHz, is positioned at the center of
the floor and connected via coaxial cable to a receiver with electromagnetic 
shield outside the anechoic chamber. The whole experiment is implemented 
by following three steps:

% fig.1
\begin{figure}
\centering
\includegraphics[width=10cm, angle=0]{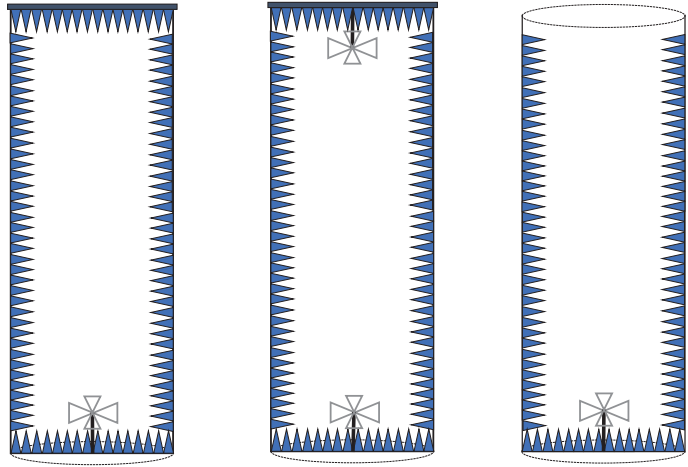}
 \caption{A schematic of three-position switch measurements in a
deep anechoic chamber} 
\label{Fig:plot1}
\end{figure}

Firstly, we conduct a noise radiation measurement by closing the ceiling 
to create a radio-quiet space inside the anechoic chamber, which is 
completely free of electromagnetic radiation and reflection. The readout
of the receiver in this case gives the system noise temperature at time $t_1$ 
%1
\begin{equation}
T(t_1)=T_{\rm sys}(t_1).
\end{equation}

Secondly, we suspend a linearly polarized cross dipole  
on the ceiling, which serves as a radiometric calibration source 
for the receiver dipole element on the floor. This calibrator system 
including signal generator, transmission cable and dipole antanna 
should be tested and calibrated in lab before shipping. 
Except the dipole on the ceiling, the whole system is 
placed outside the anechoic chamber and electromagnetically shielded. 
If the vertical shaft is 
depth enough to meet the far-field condition, the incident radiation on
the receiver antenna can be treated as a plane wave.      
When the calibrator is switched on, the output of the receiver can be 
described by a beam averaged surface brightness temperature at the frequency 
$\nu$ and time $t_2$
%2
\begin{equation}
T(\nu,t_2)=T_{\rm sys}(t_2)+G(\nu,t_2)\left(1-|\Gamma(\nu,t_2)|^2\right)
         \frac{
          \int_{\Omega_s} A(\theta,\phi,\nu)T_{\rm cal}(\theta,\phi,\nu,t_2)d\Omega}
          {\int A(\theta,\phi,\nu)d\Omega},
\end{equation} 
where $G$ is the gain of the receiver, 
$\Gamma$ is the reflection coefficient accounting for the impedance mismatch 
between the receiver antenna and the receiver itself (Rogers \& Bowman 
\cite{Rogers12}; Price et al. \cite{Price18}), 
$A$ is the receiver antenna beam pattern, and
$T_{\rm cal}$ represents the surface brightness temperature of the calibrator. 
The integration in the numerator is performed over the solid angle 
$\Omega_s$ subtended by the shaft ceiling seen at the receiver antenna, 
while the integration in the denominator is performed over the main beam 
of the receiver antennas. Now we introduce the average temperature of 
the calibrator $\bar{T}_{\rm cal}$ by taking the $T_{\rm cal}$ term out 
of the integral, yielding 
%3
\begin{equation}
T(\nu,t_2)=T_{\rm sys}(t_2)+G(\nu,t_2)\left(1-|\Gamma(\nu,t_2)|^2\right)
          \bar{T}_{\rm cal}(\nu,t_2)
         \frac{\int_{\Omega_s} A(\theta,\phi,\nu)d\Omega}
         {\int A(\theta,\phi,\nu)d\Omega}.
\end{equation} 

Thirdly, we open the cover ceiling to allow the sky radiation to enter into 
the anechoic chamber. The beam averaged surface brightness temperature 
$T_{\rm sky}$ now reads
%4
\begin{equation}
T(\nu,t_3)=T_{\rm sys}(t_3)+G(\nu,t_3)\left(1-|\Gamma(\nu,t_3)|^2\right)
          \frac{
          \int_{\Omega_s} A(\theta,\phi,\nu)
               T_{\rm sky}(\theta,\phi,\nu,t_3)d\Omega}
          {\int A(\theta,\phi,\nu)d\Omega}.
\end{equation} 
Similarly, we can use the average sky temperature $\bar{T}_{\rm sky}$ to replace
$T_{\rm sky}(\theta,\phi,\nu,t_3)$ such that 
%5
\begin{equation}
T(\nu,t_3)=T_{\rm sys}(t_3)+G(\nu,t_3)\left(1-|\Gamma(\nu,t_3)|^2\right)
              \bar{T}_{\rm sky}(\nu,t_3)
              \frac{
          \int_{\Omega_s} A(\theta,\phi,\nu)d\Omega}
          {\int A(\theta,\phi,\nu)d\Omega}.
\end{equation} 
Because the sky radiation can evidently be treated as plane wave, the above
integral is performed over the same solid angle as the one in Eq.(3).

We now work with the following ratio
%6
\begin{equation}
\frac{T(\nu,t_3)-T_{\rm sys}(t_3)}{T(\nu,t_2)-T_{\rm sys}(t_2)}
 =\frac{G(\nu,t_3)\left(1-|\Gamma(\nu,t_3)|^2\right)
      \bar{T}_{\rm sky}(\nu,t_3)}
  {G(\nu,t_2)\left(1-|\Gamma(\nu,t_2)|^2\right)
   \bar{T}_{\rm cal}(\nu,t_2)}.
\end{equation} 
If the three-position switch measurements are implemented within 
a short time so that the receiver system characterized by 
the receiver temperature $T_{\rm sys}(t)$, the gain $G(\nu,t)$  and 
the reflection coefficient $\Gamma(\nu,t)$ remain unchanged, 
we can get the average sky temperature at frequency $\nu$ and time $t$ 
through
%7
\begin{equation}
\bar{T}_{\rm sky}=\frac{T_3-T_{\rm sys}}{T_2-T_{\rm sys}}\bar{T}_{\rm cal}.
\end{equation}
The sky temperature can be decomposed into foreground 
contribution ($T_{\rm f}$) including the Galaxy and extragalactic sources and 
background one ($T_{\rm 21cm}$) from EoR---the signal
that we expect to detect in the experiment. For a featureless-spectrum 
foreground dominated by synchrotron radiation, it is possible to find 
the best-fitted foreground $\bar{T}_{\rm f}^{\rm fit}$ from Eq.(7), 
following the mature algorithms developed 
in past two decades from the simple low-order polynomial fit 
(e.g. Wang et al. \cite{Wang06}; Pritchard \& Loeb \cite{Pritchard10})
to various sophisticated techniques such as the Markov Chain Monte 
Carlo approach (Harker et al. \cite{Harker12}), 
the independent component analysis (Chapman et al. \cite{Chapman12}), 
and the Bayesian techniques (Bernardi et al. \cite{Bernardi16}). 
This allows us to extract the EoR signal simply from 
%8
\begin{equation}
\bar{T}_{\rm 21cm}=\bar{T}_{\rm sky} - \bar{T}_{\rm f}^{\rm fit}.
\end{equation}
Note that unlike all the ongoing and planned experiments which need precise 
knowledge of antenna beam pattern and its spatial and spectral variation, 
the above measurement of $T_{\rm 21cm}$ is entirely independent of 
antenna property, and the external calibrator $T_{\rm cal}$ turns to be 
the key component for the success of such experiment. 

Now we study the feasibility of the above design by examining the sensitivity 
of the system represented by the RMS variation in surface brightness temperature 
%9
\begin{equation}
\Delta T_{\rm b}=\frac{\lambda^2T_{\rm sys}}{A_e\Omega_s\sqrt{\Delta t\Delta\nu}},
\end{equation}
where $A_e$ is the effective area of the antenna, $\Delta t$ is 
the integration time, and  $\Delta\nu$ is the bandwidth. The solid angle   
$\Omega_s$ subtended by the shaft ceiling is 
%10
\begin{equation}
\Omega_s=2\pi\left(1-\frac{D}{\sqrt{D^2+R^2}} \right),
\end{equation}
in which $R$ and $D$ represent the radius and depth of the shaft, respectively. 
If the system noise is dominated by the Milky Way and follows 
a power-law of $T_{\rm sys}=60{\rm K}(\nu/300{\rm MHz})^{-2.55}$, 
the sensitivity can be quantitatively estimated through
%11
\begin{equation}
\Delta T_{\rm b}=4.82{\rm mK}\frac{1}{f(R,D)}
               \left(\frac{\nu}{100{\rm MHz}}\right)^{-4.55}
               \left( \frac{A_e}{\rm m^2} \right)^{-1}
               \left( \frac{\Delta t}{\rm day} 
                      \frac{\Delta\nu}{\rm MHz} \right)^{-1/2},
\end{equation}
where $f(R,D)=1-D/\sqrt{D^2+R^2}$ is the geometrical factor. For the
experiment on the ground, $f(R,D=0)=1$ and $\Omega_s=2\pi$. 
In this case one can easily 
achieve a sensitivity of a few mK at 100 MHz within a day. However, 
for our experiment inside an underground anechoic chamber, the calibrator 
antenna is assumed to share approximately the same patch of the sky 
as that of the cosmological signal.  This requires that 
the depth ($D$) of the shaft should be made to be much larger than 
its diameter ($2R$). Taking the more relaxed case of $D=2\times 2R$ 
as an example, we have $f(R,D=2R)\approx0.03$ 
and $\Omega_s\approx3\%\times2\pi$. To reach a 
detection limit of $\Delta T_{\rm b}=10$ mK at $\nu=100$ MHz with an 
effective area of $A_e=1$ m$^{2}$ and a bandwidth of $\Delta\nu= 1$ MHz, 
one has to integrate the measurements at open-ceiling position for 261 days. 
Considering the fact that the real measurements is carried out in a way of
three-position switch and the efficiency of operation is apparently much 
less than $100\%$,  we may have to implement the experiment on a time scale 
of up to 5--10 years in order to detect the global EoR signal. The main reason
behind the long integration time is the small sky coverage  $\Omega_s$, 
subtended by the shaft window, which prevents the antenna from receiving 
sky radiation from all directions---the key for capture of the global 
signature. 

\section{Indoor calibration: a shallow anechoic chamber} 
\label{sect: indoor2}

If the main conceptual design 
based on underground anechoic chamber in the above section
is adopted, the only way to enhance the sensitivity is to increase the field
angle $\Omega_s$ so that the antenna placed on the floor can observe a large 
patch of the sky. A shallow shaft design certainly meets the purpose, which
however, breaks the plane-wave assumption in the case of deep shaft. 
In particular, 
a dipole antenna on the ceiling can no longer serve as a calibrator 
because its radiation field does not mimic the sky signals anymore. 
So, the adoption of a shallow shaft design means that a new calibration
approach should be developed.

We begin with the geometric parameters of the shaft within 
the underground anechoic chamber to make sure that the sensitivity 
of the receiving dipole antenna on the floor can  reach a detection 
limit of $\sim 10$ mK within a reasonable integration time, say less
than a few months. While there are numerous combinations in (R, D) space 
to play with, here we focus on a simple choice 
of $R=4$ m and $D=3$ m to demonstrate the design. Such an illustrative model
yields $f(R=4,D=3)=2/5$ and  $\Omega_s=4\pi/5$, covering $40\%$ of the 
sky above the horizon. Correspondingly, the sensitivity reads
%12
\begin{equation}
\Delta T_{\rm b}=12.1{\rm mK}
               \left(\frac{\nu}{100{\rm MHz}}\right)^{-4.55}
               \left( \frac{A_e}{\rm m^2} \right)^{-1}
               \left( \frac{\Delta t}{\rm day} 
                      \frac{\Delta\nu}{\rm MHz} \right)^{1/2}.
\end{equation}
A detection limit of 10 mK is achievable within a few days 
over a broad frequency range except at low frequency end. For example, 
one may has to accumulate the data 
at the open-sky mode for about 20 days at $\nu=75$ MHz. This, however, 
does not affect the model for the purpose of illustration.

% fig.2
\begin{figure}
\centering
\includegraphics[width=10cm, angle=0]{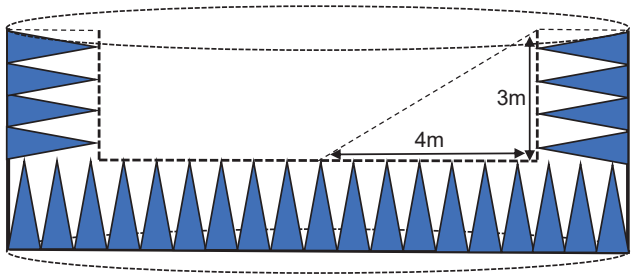}
 \caption{An illustrative model of the shallow anechoic chamber in the 
open-ceiling mode.} 
\label{Fig:plot2}
\end{figure}

The real challenge is the design of a different type of calibrator 
to replace the dipole antenna suspending on the ceiling in the case 
of deep shaft. 
Yet, the radiation field of the new calibrator should exhibit the same 
behaviour as that of the sky over the solid angle $\Omega_s$. 
Although there are probably other solutions to the problem,    
here we only explore two technical schemes to meet the requirement. 

The first choice is the drone-based calibrator, which has been 
successfully applied in the beam calibration of radio telescope
(Chang et al. \cite{Chang15}). In particular, 
this external calibration method has also been tested in 
the low frequency antennas including SKALA designed for SKA1-low
(Virone et al. \cite{Virone14})
and low frequency array dedicated to the detection of EoR signal 
(e.g. Jacobs et al. \cite{Jacobs16}). While technically the method has become  
mature and reliable, the employment of the drone-mounted 
calibrator in our experiment is unfortunately impractical. An immediate
reason is the violation of our indoor calibration principle: 
Radiation from the calibrator is contaminated by background sky when  
the underground anechoic chamber is exposed to ambient environment even if
there is no RFI in experimental site. 
A more serious problem is actually the flight time to smoothly cover the 
field of view, $\Omega_s$. Although we could carefully design the flight 
pattern in terms of either rectangular grids or concentric tracks above 
the antenna to have a smooth coverage of $\Omega_s$, it takes rather 
a long time to complete the calibration process, during which 
there is no guarantee that the receiver system remains stable. Recall 
that the receiver properties characterized by  $T_{\rm sys}(t)$, $G(\nu,t)$  and 
$\Gamma(\nu,t)$ are assumed unchanged in the three-position 
switch measurements [see Eq.(6)]. Therefore, we will not explore further 
the calibration with drone unless there is an independent way to measure
the system temperature by adding an internal noise source.

The second approach is to place a spherical dish antenna as a new ceiling 
to fully cover the anechoic chamber. The spherical dish itself will act  
as a calibrator because it can generate an isotropic radiation field, 
in a similar way as the sky signal, toward the dipole antenna on the floor.  
Our last two-step measurements will then switch between this artificial 
radio `sky' and the open sky positions, both of which subtend 
the same solid angle  $\Omega_s$ seen at the dipole antenna at the floor.  
This requires that the spherical dish should be designed to be removed or
opened conveniently and quickly in the three-position switch measurements. 
Taking the illustrative model of $R=4$ m and $D=3$ m as an example, we need 
a spherical dish antenna of 8 m in diameter with a curvature radius of 5 m 
to the focus point, which is exactly the position of the receiving dipole 
on the floor. So, the feed of the spherical dish should be mounted 
at the position of the receiving dipole, and the transmitting 
and reflecting signal should exhibit a spectral flatness over 50--200 MHz.
Although one could use the same dipole for both transmitting and 
receiving purpose, the reflecting signal from the dish surface as a calibrator 
deviates from an isotropic radiation field. One could further explore the 
possibility of designing a compact feed of spherical beam symmetry over  
$\Omega_s$ in a low frequency wideband of 50--200 MHz such as the 
sinuous antenna tested in the HERA experiment (e.g. Garza et al.
 \cite{Garza18}). However, it is mechanically impractical to put both the 
feed and the receiving dipole antenna exactly at the same position. 

Yet, there is a third way to generate an isotropic radiation field over 
$\Omega_s$ by combining and modifying the design concepts of 
the first and second method above: 
Instead of utilization of the feed at the focus point for the 
spherical dish, we can suspend a movable, dual polarized dipole antenna 
as a calibrator 
on the surface of the spherical cap. The track of the moving calibrator
can be designed to maximally and smoothly cover the dish surface 
though the calibration is 
actually performed at finite and discrete grids over a certain time. 
A continuous calibration process is also possible if the speed of the
calibrator can be precisely controlled to be constant.  
Another fundamental change is that the dish should be made of 
non-metal material to absorb radiation or eliminate reflection
from the calibrator. Furthermore, the dish surface should not be covered with 
wedge-shaped absorbers to allow the calibrator antenna to move freely. 
It is easy to show that in this case our illustrating model of $R=4$ m 
and $D=3$ m satisfies the far-field requirement: 
the distance (5 m) of the calibrator to the receiving 
antenna is larger than twice of the square of the antenna size 
($\sim 1$ m) over the shortest wavelength (1.5 m). 
While such a design seems feasible and operational, there are still two
major concerns for the measurement of the global EoR signal: 
First, the time to complete the whole calibration process may be too 
long to maintain the stability of the receiver system---a key assumption 
to remove the receiver effect in the three-position switch. 
Second, the calibration can only be performed on finite grids or tracks 
on the disk surface, and it is never possible to do the calibration at 
every position of the spherical cap. Mathematical modeling should be invoked 
to quantity the calibration error from the discrete sampling. 
Therefore, the spherical dish cover with a movable calibrator antenna 
is still not an ideal and practical design for our purpose.

\section{Outdoor calibration: diurnal motion} 
\label{sect: outdoor1}

We wish to keep the basic design (e.g. the illustrative model)
of the underground shallow anechoic 
chamber in the above section to benefit from both radiation-free environment  
and high sensitivity, and explore the possibility of calibrating the 
system using outdoor sources. In this case, the anechoic chamber will work 
in a complete open-ceiling mode while the ceiling is nevertheless needed 
to create a radiation-free space to measure the system noise 
temperature $T_{\rm sys}$, the so-called first step in the experiment. 
Man-made calibrators such as the drone-based one are difficult to meet
our purpose due to the limitation of uniformity and smoothness of 
sky coverage and lower efficiency as discussed in Section 3. 
We should therefore rely on celestial sources in the low-frequency sky 
to fulfill the task. In particular, because the celestial calibrators should 
share the same sky and frequency coverage as those of the EoR signal, 
very bright radio sources such as the Cas A or the Moon 
suggested in literature (e.g. Shaver et al. \cite{Shaver99}) 
cannot be chosen. 
The candidate is therefore the Milky Way, the brightest diffuse foreground, 
or extragalactic radio sources distributed isotropically in the sky. 

The first step remains essentially the same as that in the indoor calibration 
discussed the above,  
and the system noise $T(t_1)=T_{\rm sys}(t_1)$ is measured within 
the anechoic chamber at time $t_1$. The main difference arises from 
following steps: The anechoic chamber will receive radiation from 
the sky in an open-ceiling mode in the two-position switch between 
calibration (second step) and data acquisition (third step). Actually, 
we will not make a distinction between the calibration and measurement 
processes any more, which are carried out exactly in the same way 
except at different time. 
Denoting the possible sky variation between the latter two-position switch 
as  $\Delta T$, we get the system response at time $t_2$
%13
\begin{equation}
T(\nu,t_2)=T_{\rm sys}(t_2)+G(\nu,t_2)\left(1-|\Gamma(\nu,t_2)|^2\right)
         \frac{
          \int_{\Omega_s} A(\theta,\phi,\nu)
      [T_{\rm sky}(\theta,\phi,\nu,t_2)-\Delta T(\theta,\phi,\nu,t_2)/2]d\Omega}
          {\int A(\theta,\phi,\nu)d\Omega}.
\end{equation} 
Introducing the average sky temperature $\bar{T}_{\rm sky}$ and its variation 
$\Delta\bar{T}$ over the solid angle $\Omega_s$, 
we rewrite the above equation as
%14
\begin{equation}
T(\nu,t_2)=T_{\rm sys}(t_2)+G(\nu,t_2)\left(1-|\Gamma(\nu,t_2)|^2\right)
          \left[\bar{T}_{\rm sky}(\nu,t_2)-\Delta\bar{T}(\nu,t_2)/2 \right]
         \frac{\int_{\Omega_s} A(\theta,\phi,\nu)d\Omega}
         {\int A(\theta,\phi,\nu)d\Omega}.
\end{equation} 
Similarly, the system readout at time $t_3$ is 
%15
\begin{equation}
T(\nu,t_3)=T_{\rm sys}(t_3)+G(\nu,t_3)\left(1-|\Gamma(\nu,t_3)|^2\right)
          \left[\bar{T}_{\rm sky}(\nu,t_3)+\Delta\bar{T}(\nu,t_3)/2\right]
         \frac{\int_{\Omega_s} A(\theta,\phi,\nu)d\Omega}
         {\int A(\theta,\phi,\nu)d\Omega}.
\end{equation} 
We now work with the following ratio
%16
\begin{equation}
\frac{T(\nu,t_2)-T_{\rm sys}(t_2)}{T(\nu,t_3)-T_{\rm sys}(t_3)}
 =\frac{G(\nu,t_2)\left(1-|\Gamma(\nu,t_2)|^2\right)
      [\bar{T}_{\rm sky}(\nu,t_2)-\Delta\bar{T}(\nu,t_2)/2] }
  {G(\nu,t_3)\left(1-|\Gamma(\nu,t_3)|^2\right)
      [\bar{T}_{\rm sky}(\nu,t_3)+\Delta\bar{T}(\nu,t_3)/2] }.
\end{equation} 
Again, if the system remains stable throughout the three-position switch  
measurements, the above equation reduces to  
%17
\begin{equation}
\frac{T_2(\nu,t)-T_{\rm sys}(t)}{T_3(\nu,t)-T_{\rm sys}(t)}
 =\frac{\bar{T}_{\rm sky}(\nu,t)-\Delta\bar{T}(\nu,t)/2}
  {\bar{T}_{\rm sky}(\nu,t)+\Delta\bar{T}(\nu,t)/2 }.
\end{equation} 
Most importantly, it appears that the method would fail if 
$\Delta\bar{T}(\nu,t)=0$, indicating that it is the time-varying component 
that acts as the calibrator rather than the total foreground radiation. 
The diurnal motion of the Galactic signal has thus been suggested and 
tested as a calibrator in the EoR experiment to separate out the EoR signal 
by EDGES (Memo \#048, \#055, \#202 \& \#215 \footnote
{http://www.haystack.mit.edu/ast/arrays/Edges/EDGES-memos/}). 
However, this method has not been
adopted by EDGES mainly due to its sensitive dependence on beam correction. 
SCI-HI (Voytek et al. \cite{Voytek14}) 
also applied the 24 hour data of the Galactic 
variation to calibrate the total power spectrum. Here we explore further 
the possibility because our method is entirely unaffected by beam pattern.
The new concern is, however, whether the variation of the Galactic noise 
is actually visible/observable 
during the latter two position switching at any time of a day. 

Figure 3 shows an example of the diurnal variation of the Milky Way 
in 1 minute interval over 24 hours at 100 MHz, 
in which an imaginary observation 
is made on January 1, 2021 at the location of the 21CMA site 
(longitude 68.68$^{\circ}$, latitude 42.93$^{\circ}$) 
(Huang et al. \cite{Huang16}; Zheng et al. \cite{Zheng16}) and a Gaussian 
beam of FWHM=43.3$^{\circ}$ is assumed for the receiving antenna. 
Our demonstration is based on the diffuse Galactic radio
emission model proposed by 
de Oliveira-Costa et al. (\cite{deOliveiraCosta08}). 
It appears that the amplitude of the sky-averaged brightness
temperature of the Milky Way can reach $\sim1000$ K even within 1 minute, 
suggesting that the diurnal motion of the Galactic signal could  
be used as an ideal calibrator in terms of magnitude alone 
for the EoR experiment.

% fig.3
\begin{figure}
\centering
\includegraphics[width=10cm, angle=0]{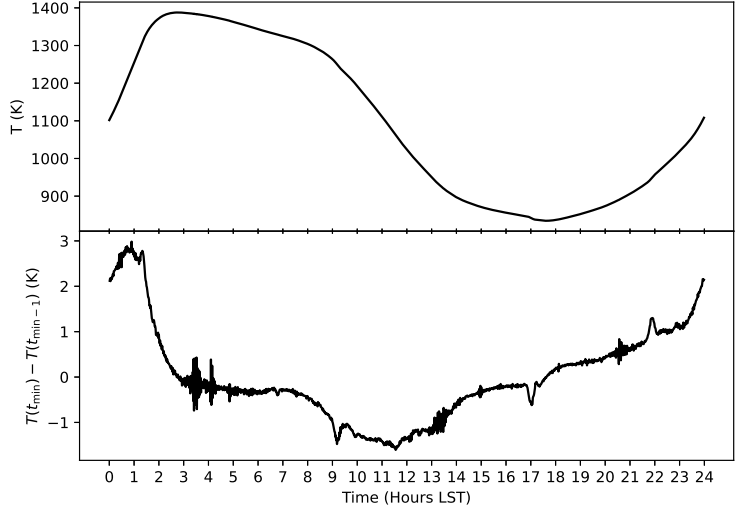}
 \caption{Diurnal variation (top panel) and difference (bottom panel) of 
the Galactic diffuse emission in 1 minute interval over 24 hours at 100 MHz, 
predicted by the simulations of de Oliveira-Costa et al. 
(\cite{deOliveiraCosta08}).    
The 21CMA site and an imaginary observation on January 1, 2021 are taken  
for this illustration. }
\label{Fig:plot3}
\end{figure}

Denoting the sky temperature  $\bar{T}_{\rm sky}$ as the foreground 
component $\bar{T}_f$ plus the EoR signal $T_{\rm 21cm}$ and keeping the 
ratio of $T_{\rm 21cm}/\bar{T}_f$ to the first order, we can rewrite Eq.(17) as 
%18
\begin{eqnarray}
\frac{T_3-T_2}{T_3-T_{\rm sys}} 
 & = & \frac{\Delta\bar{T}}{\bar{T}_f}
 \left\{ 1-\frac{1}{2}\frac{\Delta\bar{T}}{\bar{T}_f}
        +\frac{1}{4}\left(\frac{\Delta\bar{T}}{\bar{T}_f}\right)^2
        +O\left(\frac{\Delta\bar{T}}{\bar{T}_f}\right)^3  \right. \nonumber\\
 &   & \left.   -\frac{T_{21cm}}{\bar{T}_f}
        \left[1 - \frac{\Delta\bar{T}}{\bar{T}_f}
        +\frac{3}{4}\left(\frac{\Delta\bar{T}}{\bar{T}_f}\right)^2
        +O\left(\frac{\Delta\bar{T}}{\bar{T}_f}\right)^3
        \right]
 \right\}.
\end{eqnarray} 
In a logarithmic expression, the above equation takes the form of 
%19
\begin{equation}
\ln\frac{T_3-T_2}{T_3-T_{\rm sys}} =
  \ln\left(\frac{\Delta\bar{T}}{\bar{T}_f}\right)
   +{\cal F} \left(\frac{\Delta\bar{T}}{\bar{T}_f}\right)
   -\frac{T_{21cm}}{\bar{T}_f},
\end{equation}
where $\cal F$ is an analytical function of 
$\Delta\bar{T}/\bar{T}_f$. Subtracting a best-fitted smooth component or
a polynomial in frequency domain would allow us to find the EoR signal 
$T_{21cm}$ calibrated by the sky averaged foreground brightness 
temperature $\bar{T}_f$ (rather than the variation $\Delta\bar{T}$). 
Yet, application of this method 
depends critically on an accurate understanding and modelling
of the diurnal motion of the Galactic diffuse radiation 
in the frequency range of 50--200 MHz.
The typical accuracy in current models is between 5\% and 15\% 
across the entire sky (de Oliveira-Costa et al. \cite{deOliveiraCosta08}; 
Zheng et al. \cite{Zheng17}). The error in the global EoR signal resulting 
from this inaccuracy can be estimated from Eq.(19) 
%20
\begin{equation}
\frac{dT_{21cm}}{T_{21cm}} = \frac{d\bar{T}_f}{\bar{T}_f} +
                       O(1)\frac{d\Delta\bar{T}_f}{\Delta\bar{T}_f}    
                            \approx 2 \frac{d\bar{T}_f}{\bar{T}_f},
\end{equation}
in which the diurnal motion term shares the same accuracy as the Galactic 
emission one. Therefore, the error in ${T}_{21cm}$ from current modelling 
of the Galactic diffuse radiation is less than $30\%$.

In order to examine whether this algorithm allows us to efficiently  
extract the EoR signal, we simulate a set of measurements of up to 72 hours 
with the observational parameters in Fig.3 and the setup in Fig.2. 
For the 21 cm global background, we adopt a phenomenal model  
of Gaussian absorption profile (Bowman et al. \cite{Bowman18})
%21
\begin{equation}
T_{21cm}=a\exp\left[-\frac{(\nu-\nu_{21})^2}{2\sigma^2}\right],
\end{equation}
where $a=-150$ mK, $\nu_{21}=78.3$ MHz and $\sigma=5$ are the amplitude, 
peak position and deviation of the absorption trough, respectively.
We also run a flattened Gaussian profile for comparison:

%22
\begin{equation}
T_{21cm}=a\left\{\frac{1-\exp[-\tau\exp(b)]}{1-\exp(-\tau)}\right\},
\end{equation}
in which 
%23
\begin{equation}
b=\frac{4(\nu-\nu_0)^2}{w^2}\log
  \left[-\frac{1}{\tau}\log\left(\frac{1+\exp(-\tau)}{2}\right) \right].
\end{equation}
All the parameters $a$, $\nu_0$, $w$ and $\tau$ can be fixed by fitting 
the observed absorption trough from EDGES \cite{Bowman18}:  
$a = -520$ mK, $\nu_0 = 78.3$ MHz, $w=20.7$ MHz, and $\tau = 7$.

In the mock observation, three-position switch is performed in a time interval 
of 1 minute each with bandwidth of 50 kHz, 
namely, a total of 3000 sampling points 
over 150 MHz ($50-200$ MHz) bandpass are taken. We assume a constant receiving 
system temperature of $T_{\rm sys} = 50$ K. A series of `observing' data 
($T^i_1=T^i_{\rm sys}=50$ K, $T^i_2, T^i_3$; $i=1,2,...,3000$) would be 
collected every three minutes. For each set of measurement (3 minutes)   
we find the best-fit 
polynomial in frequency domain from $\ln[(T^i_3 - T^i_2)/(T^i_3 - T^i_{\rm sys})]$ 
and then obtain the residual $\Delta^j$. We accumulate and average 
the residuals over a sufficiently long time to reach the desired sensitivity. 
Eventually, the 21cm global signals can be extracted through
$\bar{T}_{\rm 21cm}= -\bar{T}_f \bar{\Delta}$. 
Fig.4 shows the low-frequency sky model,  
$T_{\rm sky}= T_{\rm sys}+\bar{T}_{\rm 21cm} + \bar{T}_f$, 
two input models of the cosmic EoR background $\bar{T}_{\rm 21cm}$ 
described by Eqs.(21) and (22), respectively, and 
the recovered signals, in which mock observations over 72 hours 
or 1440 sets of measurements are simulated. 
 
%fig.4
\begin{figure}
    \centering
    \includegraphics[width=0.8\textwidth]{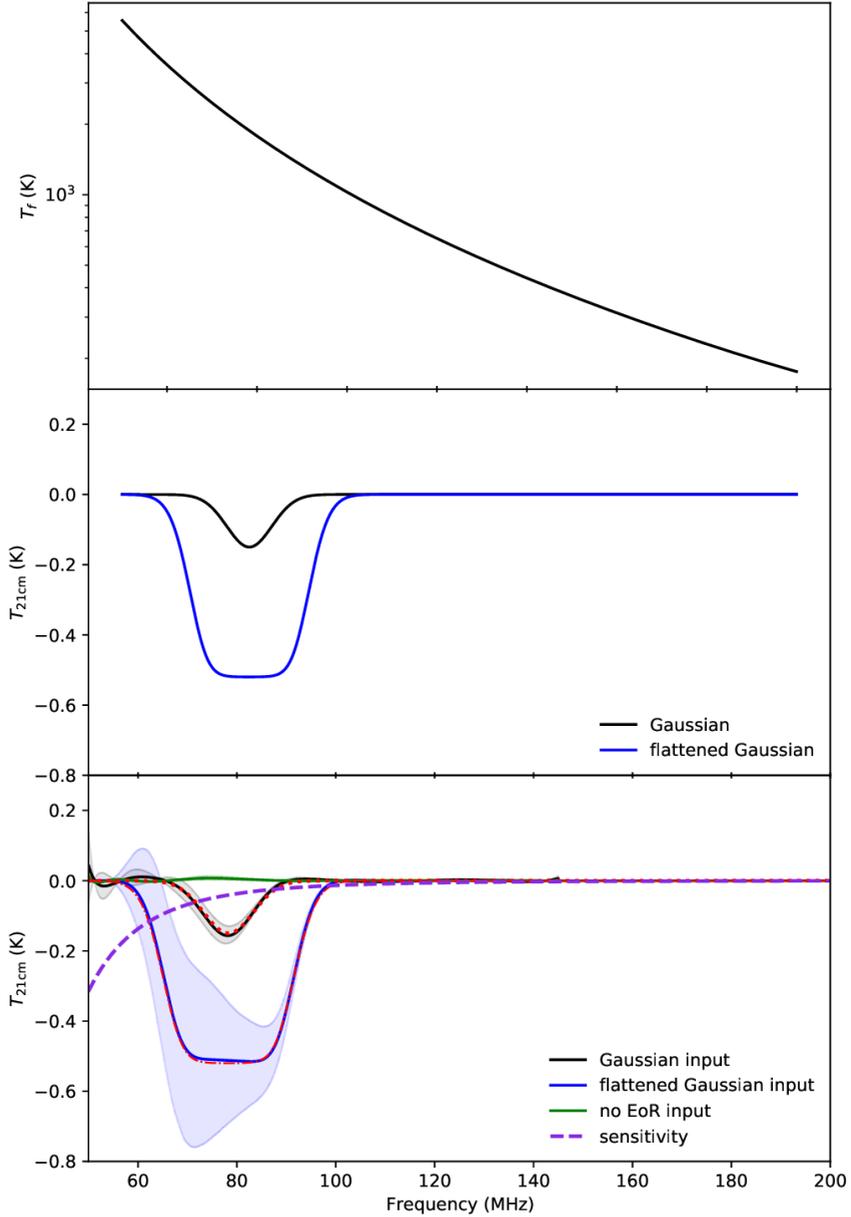}
    \caption{
\textbf{Top panel}: The input Galactic diffuse emission model  
(\cite{deOliveiraCosta08}) as a function of observing frequency.  
\textbf{Middle panel}: The input 21cm EoR signal   
described by a Gaussian absorption profile (black lines) 
and a flattened Gaussian absorption profile (blue line), respectively. 
\textbf{Bottom panel}: The recovered 21cm global signals from 72 hours 
(1440 sets of measurements) of mock observations. The shadow regions in each 
case indicate the dispersions represented by 14--86 percentile 
of the recovered signal around 
the average (solid line) over 1440 measurements. Overlaid 
dotted/dot-dashed lines are the corresponding original input signals. 
The large dispersions arise partially from the 'calibrator' $T_f$, which 
demonstrates a diurnal variation as shown in Fig.3. For comparison, 
we have also shown the result for the Galactic emission alone as the 
input signal using the same algorithm of signal extraction. 
Dashed line indicates the predicted sensitivity based on 
a bandwidth of $\Delta\nu= 1$ MHz for an integration time of
24 hours, in which an efficiency of $1/3$ has been used.
}
    \label{Fig:Plot4}
\end{figure}

Technically, we first employ the Savitzky-Golay filter with a window of 
$\sim 10$ MHz in the fitting of 
$\ln[(T^i_3 - T^i_2)/(T^i_3 - T^i_{\rm sys})]$ over 50--200 MHz. 
It turns out that the residual errors are within an acceptable level 
except in a certain frequency range where the radiation is mixed with 
the background EoR signal, reflected by some relatively large dispersions 
and oscillations. 
We then perform the second polynomial fitting by blanking the data in 
the `noise' region. The residual between the measurement and
the newly fitted polynomial will be accumulated to extract the EoR signal. 

Finally, one may argue that the Galactic diffuse emission and 
its diurnal motion demonstrate complex spatial structures and time variation, 
while the 21 cm radiation from EoR is isotropic in the sky and remains almost 
time-invariant. Apparently, the external calibrator no longer 
mimics the background radiation.  For this reason, if the measurements are made 
during the Galaxy down, the effect of the Galactic structures may be largely 
reduced. On the other hand, if the spatial response of the receiving antenna, 
$A(\theta,\phi,\nu)$, is designed to be a smooth function of observing 
frequency, the above algorithm can still be used to extract the EoR signal, as 
demonstrated by our mock observation.

\section{Outdoor calibration: polarization} 
\label{sect: outdoor2}

Measurement of polarization plays a key role in modeling and subtracting 
foreground contamination in EoR experiment. It has been 
realized that the low-frequency foregrounds are dominated by synchrotron 
radiation and thus polarized because of the existence of magnetic field, 
while the cosmic CD/EoR signal is thermal emission/absorption and 
therefore unpolarized. A naive approach that is 
yet worthy of further exploration is to use this unique feature as a  
calibrator in EoR experiment. The same three-position switch measurements 
can be once again employed except that latter two steps are switched between 
two polarized states. There are two ways to achieve this goal through either 
(1) a pair of perpendicularly spaced and linearly polarized dipoles or 
(2) a linearly polarized dipole mounted on a 90-degree rotating platform. 
The whole system is placed in the underground shallow anechoic 
chamber, similar to the experimental environment discussed above. 

The first design requires two perpendicularly spaced and linearly polarized 
dipoles and hence two backend receivers but the measurements 
can be made simultaneously. One can also carry out the measurement in 
two separate steps by sharing the same receiver to reduce the systematics.
Yet, for the latter the measurement would also  
contain a time-varying component from the diurnal motion of 
the Galactic emission that we have discussed in the above section. 
The calibrator $\Delta\bar{T}$ in Eq.(19) is eventually a 
combination of both the polarized and diurnal variations of the sky. 
For the former, while a simultaneous measurement of two orthogonal polarization 
components is unaffected by the diurnal motion of the Galactic emission, 
the whole experiment actually involves two independent channels 
(dipoles, receivers and DAQ). Systematics could be introduced 
even if they are fabricated to be identical. Another concern for 
both designs arises from the possible cross-talk between two 
perpendicularly spaced dipoles. 

The second design is based on a single dipole mounted on a 90-degree rotating 
platform. Three-position switch operations are also applied, 
and we only need to replace the second and third steps by measuring 
two orthogonal X--Y polarized components $T_X$ and $T_Y$ at time 
$t=t_2$ and $t=t_3$, respectively, which is achieved by rotating the
single-polarized linear dipole by 90 degrees. Although the receiver 
can be assumed to be stable and unchanged during the rotation, 
the variation of the sky brightness temperature in such a measurement, 
however, includes both the polarized sky $\Delta\bar{T}_P=T_Y-T_X$ 
and diurnal motion of the 
Galactic emission $\Delta\bar{T}_{\rm MK}$.  Replacing $\Delta\bar{T}$ by 
$\Delta\bar{T}=\Delta\bar{T}_P+\Delta\bar{T}_{\rm MK}$, Eq.(19) should 
still remain valid and can be used to obtain the 21cm EoR signal, provided  
that the polarization signal does not break down the smooth spectrum    
assumption.

Unfortunately, the foreground polarization may result in a 
frequency-dependent spectrum that further exhibits a rapid and irregular 
variation as a function of Faraday depth when the linearly polarized 
radiation travels through the Galactic magnetic field 
(Rybicki \& Lightman \cite{Rybicki79}; 
Carucci et al. \cite{Carucci20}). 
The exact magnitude of this polarization
leakage into the EoR signal is still uncertain   
(e.g. Moore et al. \cite{Moore13}, \cite{Moore17}; 
Asad et al. \cite{Asad15}). 
Recent simulation involving a more realistic model of Galactic polarized 
synchrotron emission and a single polarization antenna suggested 
that the polarized foreground not only can give rise to a complex 
frequency structure but also produce an enhanced and distorted 
21 cm absorption though similar to the anomalous profile detected by
the EDGES experiment (Spinelli et al. \cite{Spinelli18}).   
Before further investigations are needed to clarify the issue, 
for the time being we would not recommend the foreground polarization 
as an external calibrator in the global EoR experiment.

\section{Discussion and conclusions} 
\label{sect:discussion}

We have presented a conceptual design study of four types of 
external calibrators for the 21 cm EoR experiment aiming at measuring 
the globally averaged sky brightness in frequency range of 
50--200 MHz. Unlike the internal calibrators widely adopted in 
current EoR experiments, external calibrator seeks to mimic 
a radiation field similar to that of the EoR signal. This allows 
one to completely remove instrumental effect such as direction- and
frequency-dependent beam of the antenna, frequency-dependent gain of 
the receiver, and even the unstability of the system if measurements
are implemented within a short time interval. The whole system is placed  
in an underground shallow anechoic chamber with an open/close
ceiling to further reduce 
the environmental effect such as RFI and ground radiation/reflection.
Conventional three-position switch measurements are implemented among 
system noise, external calibrator and low-frequency sky.  
 
We have explored two external calibrators for each of the indoor and outdoor 
calibration. It appears that two types of calibrator in the indoor calibration 
based on artificial emitting sources fail 
to meet our purpose, due to either small sky coverage or difficulties of
engineering realization.  Outdoor calibration relies on celestial astronomical 
sources in low frequencies. The polarized foreground especially the 
Milky Way as a calibrator is not recommended because the Faraday rotation 
of the Galactic magnetic would break down the smooth spectrum scenario, 
a key to remove the foreground contamination, though it is still uncertain 
whether the magnitude of such a polarization leakage is comparable to 
that of the EoR signal. Therefore, the possible candidate in the outdoor 
calibration turns to be the diurnal motion of the Galactic diffuse emission. 
Indeed, the Galactic radiation in low frequency shares the same sky coverage 
as that of the EoR signal. This provides a possibility of eliminating  
the beam effect of the receiving antenna, an advantage over 
the internal calibrator in current EoR experiments. 
While the application of this method depends on our understanding and 
modelling of the diurnal motion of the Galactic diffuse radiation 
in the frequency range of 50--200 MHz, the extraction of the EoR signal 
is actually unaffected by the exact magnitude variation $\Delta\bar{T}$ 
in radiation due to diurnal motion of the Galaxy, as shown in Eq.(19), 
provided that $\Delta\bar{T}$ does not introduce any complex spectral 
structures. Indeed, subtraction of the best-fitted foreground, say, a 
polynomial, from Eq.(19) will unveil the EoR signature calibrated by 
the total intensity of the foreground. Yet, regardless of its same  
sky coverage as that of the EoR, the Galaxy demonstrates complex spatial 
structures. To what extent the Galactic spatial structures affect 
the measurement of the EoR signal needs further investigation.  

Now it is technically feasible to construct and conduct a new experiment 
towards the detection of the global 21 cm EoR signal using 
the Galactic emission and its diurnal motion as an external 
calibrator. The antenna, receiver and data acquisition system have been
well developed by the existing experiments such as BIGHORNS, EDGES, 
LEDA, PRI$^z$M, SARAS and SCI-HI. The most expensive component in this 
experiment is perhaps the underground shallow anechoic chamber --- the 
platform to host the whole system. It would be better that the measurements 
are implemented in the 'Galaxy down' phase to maximally reduce 
the possible influence of the Galactic spatial structures. 
We have already raised sufficient funds to support such an experiment, 
and it is hoped that the design of the system can be completed recently 
and the experiment can start to collect data in about two years.

%============================================================

\begin{acknowledgements}
This work is supported by the Key Projects of Frontier Science
of Chinese Academy of Sciences under grant No. QYZDY-SSW-SLH022,
the Strategic Priority Research Program of Chinese Academy of Sciences
under grant No. XDB23000000, and the National Key R\&D Program
of China under grant No. 2018YFA0404601.
\end{acknowledgements}

\label{lastpage}

\end{document}